\documentclass[conference,final]{IEEEtran}

\usepackage{blindtext, graphicx}
\usepackage[colorlinks=true,linkcolor=red,urlcolor=blue,citecolor=blue]{hyperref}
\usepackage{caption}
\usepackage{subcaption}
\usepackage[colorinlistoftodos]{todonotes}
\usepackage{algorithm}
\usepackage[hang,flushmargin]{footmisc}
\usepackage{lipsum}
\usepackage{bm}
\usepackage{algpseudocode}
\usepackage{array}
\newcolumntype{P}[1]{>{\centering\arraybackslash}p{#1}}
\usepackage{booktabs}
\usepackage{float}

\ifCLASSINFOpdf
\else
\fi
\usepackage[cmex10]{amsmath}

\makeatletter
\newcommand*{\rom}[1]{\expandafter\@slowromancap\romannumeral #1@}
\makeatother

\begin{document}
%
\title{\LARGE \bf An Efficient and Scalable Controller for Cooperative Intersection Management}
\title{PAIM: \underline{P}latoon-based \underline{A}utonomous \underline{I}ntersection \underline{M}anagement}
\author{\IEEEauthorblockN{Masoud Bashiri\IEEEauthorrefmark{1},
Hassan Jafarzadeh\IEEEauthorrefmark{1} and
 Cody H. Fleming\IEEEauthorrefmark{1}}
\IEEEauthorblockA{\IEEEauthorrefmark{1} Department of Systems and Information Engineering,
University of Virginia,
Charlottesville, VA, USA}
\{\href{mailto:mb4bw@virginia.edu}{mb4bw},
\href{mailto:hj2bh@virginia.edu}{hj2bh},
\href{mailto:fleming@virginia.edu}{fleming}\}.virginia.edu}



%

\maketitle
\begin{abstract}
With the emergence of autonomous ground vehicles and the recent advancements in Intelligent Transportation Systems, Autonomous Traffic Management has garnered more and more attention. Autonomous Intersection Management (AIM), also known as Cooperative Intersection Management (CIM) is among the more challenging traffic problems that poses important questions related to safety and optimization in terms of delays, fuel consumption, emissions and reliability.
Previously we introduced two stop-sign based policies for autonomous intersection management that were compatible with platoons of autonomous vehicles. These policies outperformed regular stop-sign policy both in terms of average delay per vehicle and variance in delay.
This paper introduces a reservation-based policy that utilizes the cost functions from our previous work to derive optimal schedules for platoons of vehicles. The proposed policy guarantees safety by not allowing vehicles with conflicting turning movement to be in the conflict zone at the same time. Moreover, a greedy algorithm is designed to search through all possible schedules to pick the best that minimizes a cost function based on a trade-off between total delay and variance in delay.
A simulator software is designed to compare the results of the proposed policy in terms of average delay per vehicle and variance in delay with that of a 4-phase traffic light.
\end{abstract}

%
\IEEEpeerreviewmaketitle

\section{Introduction}
As bottlenecks of traffic flow, intersections are known to be a major contributor to traffic accidents. According to National Highway Traffic Safety Administration (NHTSA),~\cite{choi2010crash}~40 percent of all crashes that occurred in the United States in 2008 were intersection-related. Moreover, Traffic efficiency is reported to be closely correlated with traffic safety on intersections~\cite{chang2003relationship}. Traffic congestion is also among the more important contributors of~$CO_{2}$ emissions~\cite{grote2016including} which is the largest constituent of transport's greenhouse gas emissions. Vehicle stop times at intersections also contribute to carbon monoxide ($CO$) emissions. International energy agency reports that when vehicles are idle at an intersection they emit about~5--7 times as much $CO$ as vehicles traveling between~5--10 mph~\cite{statistics2011co2}.

Traffic Lights and stop signs are the major methods for traffic control  used at intersections. While traffic lights have helped improve traffic flow at intersections, they are still considered inefficient and a contributor to traffic congestion and accidents. Statistically a majority of intersection-related accidents occur in the presence of traffic lights~\cite{choi2010crash}. In the past two decades, adaptive and smart traffic light controllers have been introduced and deployed, with significant improvements in terms of delay and congestion. It has been shown that the performance of traditional traffic lights can be improved through machine learning based approaches such as Fuzzy logic~\cite{niittymaki2000signal}, neural networks~\cite{spall1997traffic} and Reinforcement Learning~\cite{abdulhai2003reinforcement}, and mathematical models such as Mixed-Integer Linear Programming (MILP)~\cite{lin2004enhanced}. However, traffic lights have remained a major contributor to congestion and traffic accidents. Moreover, while signalized intersections work well with human drivers, they don't necessarily leverage the advantages associated with autonomous vehicles.

Recent advancements of Information Technology and the emergence of Vehicular Adhoc Networks~(VANETs) that support Vehicle to Vehicle (V2V), Vehicle to Infrastructure~(V2I) and Vehicle to Pedestrian (V2P) communications have brought forth opportunities for further advancements of intersection management infrastructure. This includes new non-signalized approaches for the intersection management problem, commonly known as Cooperative Intersection Management~(CIM), where road users, i.e. vehicles communicate with the infrastructure and/or other users to cooperatively coordinate the traffic flow. In 2014, the Institute of Electrical and Electronics Engineers~(IEEE) published the Wireless Access for Vehicle Environments~(WAVE)~\cite{ieee2013ieee} specification. The standards define architectures based on Dedicated Short Range Communications~(DSRC) for which the Society of Automotive Engineers~(SAE) has specified message types and data elements through various standards.

Due to the limited capacity of current V2X communications, the communication complexity is one of the critical issues for
CIM~\cite{chen2016cooperative}. 
In~\cite{bashiri2017platoon} we proposed that platooning in the vicinity of intersections could reduce communication overhead by allowing platoon leaders to negotiate with the infrastructure and other platoons on behalf of the followers. Moreover, it can help improve the efficiency of \emph{any} scheduling policy by enabling smooth trajectories in the conflict zone. The simulation results showed that the proposed stop-sign based controller outperformed a regular stop sign by~$50\%$ in terms of average delay per vehicle and~$40\%$ in variance of delay. The proposed policies decrease computational complexity by only including one platoon per lane into the schedule.

This paper introduces a reservation-based policy that utilizes cost functions that minimize delay, or a combination of delay and variance, to derive optimal schedules for platoons of vehicles. Such schedules would decrease average delay per vehicle while decreasing the variance in delay due to the fairness of the cost functions and as a result increase intersection throughput and decrease the average fuel consumption in the vicinity of an intersection. The proposed policy guarantees safety by not allowing vehicles with conflicting turning movement to be in the conflict zone at the same time.

The rest of this paper is organized as follows. Section~\ref{sec:CIM} provides a background of previous works on the CIM problem. Section~\ref{sec:dynamics} describes the vehicle dynamics model, fuel consumption model and platooning approach used in this work. Section~\ref{sec:paim} introduces the proposed controller. In section~\ref{sec:simulations} the results of simulations of a single 4-way intersection scenario are discussed and the proposed method's results are compared with that of a 4-phase traffic light controller. Conclusions are presented in section~\ref{sec:conclusions}.
\section{Cooperative Intersection Management}{\label{sec:CIM}}
Several methods have been proposed to leverage autonomous and connected vehicles for the intersection management. The new methodologies for CIM can be categorized into two classes of Centralized and Distributed methods~\cite{rios2017survey}. In centralized methods, a central intersection manager unit receives real-time information and requests from road users and decides how to coordinate the traffic flow, e.g. instruct vehicles how, when and if to pass the intersection. Distributed methods, on the other hand, do not rely on a central control unit. Instead, all vehicles collaboratively plan their trajectories. These methods usually involve negotiation protocols to make decisions on a high level and each user/vehicle makes decisions based on shared objectives given local information from its sensors on a lower level.

\subsection{Centralized Methods}
In~\cite{dresner2004multiagent}, Dresner et al. propose a centralized resource reservation algorithm based on First Come First Serve (FCFS) policy. The control unit receives requests from all vehicles approaching the intersection, simulates the vehicle's movement through the intersection, given the information in the request, 
and confirms the request if there is no conflict with previously accepted trajectories, otherwise the request would be rejected and the vehicle has to request at a later time. The authors assume constant speed at the intersection and perform simulations comparing results with a traffic light and an overpass. It was shown that the proposed method outperformed traffic light in terms of average vehicle delay. The authors further developed their work in~\cite{dresner2005multiagent} to add several methods to improve the performance and overcome major disadvantages of the previous work. Moreover, they improved the work in~\cite{dresner2008multiagent} and~\cite{sharon2017intersection} by adding support for human drivers, considering emergency vehicles that generally have higher priority, and communication schemes for vehicles with different levels of autonomy. 

In~\cite{miculescu2014polling}, a case of two merging roads (one lane) is modeled as a
polling system with two queues and one server. The polling system determines the sequence of times
assigned to the vehicles on each lane to enter the merging road. The arrival times along with the trajectories of the leading vehicle are then used in a coordination algorithm to generate optimal trajectories for each vehicle in the queue. 

Lee and Park~\cite{lee2012development} derive a nonlinear constrained optimization problem to enhance the performance a traffic signal controller in presence of fully autonomous vehicles. A phase conflict map of the traffic signal is used as part of the optimization problem.

\subsection{Decentralized Methods}
In~\cite{azimi2012intersection} a controller model has been proposed along with different V2V-based intersection management protocols to enhance traffic throughput and safety. Each vehicle runs a collision avoidance algorithm that takes in all safety messages that are being broadcast by surrounding vehicles and detects possible collision. Estimations are then used to generate alternative trajectories to avoid collisions. 

Wu et al.~\cite{wu2015distributed} proposed a decentralized stop-and-go based algorithm that relies on wireless shared information among all approaching vehicles. The vehicle with a shorter estimated arrival time will cross the intersection, while others will need to come to a complete stop until the conflict zone is cleared. Vehicles with non-conflicting turning movements can cross simultaneously.
\subsection{Motivation and Contributions}
The existing approaches are limited with respect to at least one of the following factors: 
\begin{itemize}
    \item unrealistic or infeasible bandwidth requirements for communication
    \item no performance guarantees, i.e. no guarantee that the intersection will behave better than signalized intersections
    \item no formal safety guarantee
    \item scalability with respect to number of cars and lanes
    \item unrealistic assumptions about vehicle behavior
\end{itemize}
Any solutions for the CIM problem has to be compatible with real world communication capacity of vehicular networks, otherwise such solutions will not be feasible. To address this issue we propose a platoon-based approach where vehicles request a pass as a platoon. The intersection manager utilizes an optimization based policy to allocate slots in time and space for any platoon approaching the intersection. Communicating with platoon leaders, instead of every approaching vehicle, decreases the amount of communication needed. This approach also takes advantage of recent advances in platooning and connected vehicle control.

The performance of the solution has to be verified with respect to various performance metrics such as average delay, intersection throughput, average speed, fuel consumption, etc. We conduct microscopic simulations to verify the performance of the proposed method in comparison to a pre-timed 4-phase traffic light controller. Table~\ref{tab:metrics} shows the performance metrics used in this work.

The solution has to be scalable to growing number of roads/lanes and vehicles. We achieve a scalable solution by only taking the information from the closest platoons to the intersection at each iteration. This guarantees that the computational complexity of the algorithm remains relatively unchanged as the size of the input grows.

Most of the previously published papers make simplistic assumptions about vehicle dynamics, e.g. second order dynamics where the control input generates instant accelerations, disregarding disturbance forces, and oversimplifying the driving force, which is a function of throttle and brake positions among other factors. In this paper, we utilize a  model for vehicle dynamics to generate more realistic results when compared to the relevant literature. These assumptions help design control strategies that are more feasible for use in the real world.
\begin{table}[ht]
\vspace{-6pt}
\centering
\caption{Performance Measure Index}
\label{tab:metrics}
\begin{tabular}{@{}cc@{}}
\toprule
\multicolumn{2}{c}{\textbf{Performance Measure Index}} \\ \midrule
Performance Index                 & Unit               \\ \midrule
Average Delay                     & s                  \\ \midrule
Delay Standard Deviation                    & s                 \\ \midrule
Intersection Throughput           & veh/hour           \\ \midrule
Fuel Consumption                  & ml/veh             \\ \bottomrule
\end{tabular}
\vspace{-8pt}
\end{table}
\section{Vehicle Dynamics and Platooning}\label{sec:dynamics}
In our previous work, a simple linear model was assumed for the vehicle dynamics. Such simplistic assumptions can be found in many of the  work in Section~\ref{sec:CIM}. In what follows, 
however, a nonlinear model has been adopted and modified. The rest of this section provides the details of this model, which is divided into two parts: equation of motion and rotational velocity.

\subsection{Vehicle Dynamics}
We modify the original nonlinear dynamics found in~\cite{astrom2010feedback} in order to represent the braking and turning required to traverse an intersection.
\subsubsection{Equations of Motion}
A general form of the equation of motion is Equation~\ref{eq:em} which is a force balance for the car body.
\begin{equation}\label{eq:em}
    m\frac{dv}{dt} = F - F_d
\end{equation}
In the above equation,~$m$ is the total mass of the car,~$v$ is the car's speed,~$F$ is the traction force generated from the contact of wheels with the road and~$F_{d}$ is the sum of all disturbance forces due to gravity, friction and aerodynamic drag (Equation~\ref{eq:F_D}).
\begin{equation}\label{eq:F_D}
    F_d = F_g + F_r + F_a
\end{equation}
Equation~\ref{eq:F} models the driving force generated by the engine. The engine torque~$T$ is a function of the throttle position controlled by control input~$0 \leq u \leq 1$, and the engine speed~$\omega$ (Equation~\ref{eq:omega}).~$\alpha_{n}$ is the inverse of the effective wheel radius.
\begin{equation}\label{eq:F}
    F = \alpha_n u T(\alpha_n v) 
\end{equation}
\begin{equation}\label{eq:omega}
    \omega = \frac{n}{r}v = \alpha_n v 
\end{equation}
In the above equation,~$n$ is the gear ratio,~$r$ is the wheel radius and~$v$ is the speed of the car.
In this work, the torque is modeled by the simple representation in Equation~\ref{eq:torque}, where~$\beta$ is a design parameter of the engine and~$T_{m}$ is the maximum torque obtained from the maximum engine speed~$\omega_m$. Typical values for these parameters are,~$T_m = 190 Nm$,~$\omega_m = 420 rad/s$ and~$\beta = 0.4$.
\begin{equation}\label{eq:torque}
    T(\omega) =  T_m (1- \beta {(\frac{\omega}{\omega_m}-1)}^2)
\end{equation}\\
Equations~\ref{eq:F_G},~\ref{eq:F_R} and~\ref{eq:F_A} represent the disturbance forces due to gravity, friction and aerodynamic drag respectively.
\begin{align}
    F_g &= mgsin(\theta)\label{eq:F_G}\\
    F_r &= mgC_r sgn(v)\label{eq:F_R}\\
    F_a &= \frac{1}{2}\rho C_d Av^2\label{eq:F_A}
\end{align}
\vskip+1pt
Equation~\ref{eq:em-complete} represents the complete form of the equation of motion, where~$\alpha_n$ is the reverse of effective wheel radius, given the~$n^{\text{th}}$ gear position.~$\rho$ is the density of air,~$C_r$ is the coefficient of rolling friction,~$C_d$ is the aerodynamic drag coefficient and~$A$ is the frontal area of the car. In this work we use typical values for these parameters as follows:~$\alpha_1 = 40$,~$\alpha_2 = 25$,~$\alpha_3 = 16$,~$\alpha_4 = 12$,~$\alpha_5 = 10$,~$C_r = 0.01$,~$\rho = 1.3 \frac{kg}{m^3}$,~$C_d =~0.32$ and~$A = 2.4 m^2$. The slope of the road~$\theta$ is assumed to be zero in this work, thus eliminating the influence of the gravity as a disturbance force in the experiments.
\vskip-8pt
\begin{equation}\label{eq:em-complete}
\begin{split}
    m\frac{dv}{dt} &= \alpha_n uT(\alpha_n v) - mgC_r sgn(v)\\
    & - \frac{1}{2}\rho C_d Av^2-mg sin\theta
    \end{split}
\end{equation}

\vskip+1pt
We modify the model in~\ref{eq:em-complete} to add a braking component to the system. The braking is modeled as an extra friction component to the equation of motion. The control input~$u$ is replaced with two new control inputs~$0 \leq h \leq 1$ and~$0 \leq b \leq 1$, representing throttle and braking commands respectively, where both are functions of the computed control input~$-1 \leq u' \leq 1$ (Equations~\ref{eq:t},~\ref{eq:b}).
\vskip-6pt

\begin{align}
 h(u') &= 
  \begin{cases}
   u' & \text{if } u' > 0 \\
   0       & \text{if } u' \leq 0
  \end{cases}\label{eq:t}\\
 b(u') &= 
  \begin{cases} 
   \mid u' \mid & \text{if } u' < 0 \\
   0       & \text{if } u' \geq 0
  \end{cases}\label{eq:b}
\end{align}
\vskip+1pt
Equation~\ref{eq:em-final} shows the modified equation of motion. 
\begin{equation}\label{eq:em-final}
\begin{split}
m\frac{dv}{dt} &= \alpha_n hT(\alpha_n v) - mg(C_r + b) sgn(v)\\
&- \frac{1}{2}\rho C_d Av^2-mg sin\theta
\end{split}
\end{equation}
\vskip+2pt
\subsubsection{Rotational Velocity}
Given the model above, the rotational velocity of the vehicle can be calculated by Equation~\ref{eq:rot-vel}.
\vskip-2pt
\begin{equation}\label{eq:rot-vel}
    \frac{d\theta}{dt} = \frac{v_0}{L}tan\delta
\end{equation}
\vskip-2pt
\noindent where~$\theta$ is the vehicle's orientation,~$v_0$ is the speed of the rear wheel(s),~$L$ is the wheel base and~$\delta$ is the steering angle.
\subsection{Fuel Consumption Model}
For approximating the fuel consumption of the vehicles, here we adopt a differentiable function of velocity and acceleration introduced in~\cite{kamal2013model}, where two polynomial estimation functions of velocity and acceleration are approximated through curve-fitting process, given real-world data obtained from a typical vehicle with similar properties compared to the assumptions of this work. The fuel consumption in~$ml/s$ is estimated as:
\begin{equation}
    \label{eq:fc-1}
    f_V = (f_{cruise} + f_{accel})
\end{equation}
Where,~$f_{cruise}$ represents fuel consumption rate at a steady velocity of~$v$, and~$f_{accel}$ is the additional consumption due to presence of acceleration~$a$ at velocity~$v$.
\begin{align}
    f_{cruise} &= b_0 + b_1 v + b_2 v^2 + b_3 v^3\label{eq:fc-2}\\
    f_{accel} &= a(c_0 + c_1 v + c_2 v^2)\label{eq:fc-3}
\end{align}
The consumption parameters are~$b_0 = 0.1569$,~$b_1 = 2.450 \times 10^{-2}$,~$b_2 = −7.415 \times 10^{-4}$,~$b_3 = 5.975 \times 10^{-5}$,~$c_0 = 0.07224$,~$c_1 = 9.681 \times 10^{-2}$, and ~$c_2 = 1.075 \times 10^{-3}$. Details of the formation and determination of parameters of this fuel consumption model are described in~\cite{kamal2011ecological}.
\subsection{Cooperative Adaptive Cruise Control (CACC)}
The CACC (a.k.a platooning) problem~\cite{varaiya1993smart} has been widely studied in the literature and several solutions have been proposed~\cite{iihoshi2000vehicle,hedrick1991longitudinal,kato2002vehicle}. We adopt the Predecessor-Leader-Follower (PLF) information flow topology to describe the communication between the vehicles in a platoon.

A proportional-integral-derivative (PID) controller is designed and tuned given the vehicle dynamics to achieve string stability. An interested reader can view a video\footnote{\href{https://youtu.be/uuxMVm0MWDQ}{https://youtu.be/uuxMVm0MWDQ}} of a step response of the controller in a simple scenario where a platoon of five vehicles starts from a standstill and is supposed to reach the velocity of~$30\frac{mi}{h}$, come to a complete stop at the intersection and accelerate to~$25\frac{mi}{h}$.

\section{Platoon-based Autonomous Intersection Management}\label{sec:paim}
In this work, we extend the policies from our previous work, create an application layer communication protocol and eliminate some of the simplistic assumptions about vehicle dynamics to achieve the following goals:
\begin{enumerate}
    \item Achieve a solution under realistic assumptions
    \item Improve efficiency in terms of delay and fuel consumption
    \item Design a communication protocol that hides the algorithm from the vehicle agents
\end{enumerate}
\subsection{Communication Protocol}
A protocol is designed to ease the Vehicle to Vehicle (V2V) and Vehicle to Infrastructure (V2I) communications. Using this protocol, the intersection manager only has to communicate with the leader of a platoon. Each vehicle is broadcasting its state on a~$10Hz$ frequency, while receiving incoming packets on the same frequency. There are four types of packets designed for the platoon leaders and two types for the infrastructure as follows.
\begin{description}
  \item[Vehicle Message Types:]\ 
    \begin{itemize}
      \item \emph{Request}
      \item \emph{Change-Request}
      \item \emph{Acknowledge}
      \item \emph{Done}
    \end{itemize}
  \item[Infrastructure Message Types:]\ 
    \begin{itemize}
      \item \emph{Acknowledge}
      \item \emph{Confirm}
      \item \emph{Reject}
    \end{itemize}
\end{description}
Approaching platoon leaders send a $Request$ message once they are in a pre-defined proximity of the intersection, and await the response from the manager. Both~\emph{Request} and~\emph{Change-Request} messages consist of the unique Vehicle Identification Number (VIN) of the leader, current position, velocity and acceleration of the leader, estimated arrival time at the conflict zone and the size of the platoon, e.g. number of followers.

Upon receiving a request, the manager sends an~\emph{Acknowledge} message to the sender, runs the scheduler and responds with either an~\emph{Confirm} or a~\emph{Reject} message. The manager expects to receive an~\emph{Acknowledge} from the corresponding vehicle and if such message is not received, it re-sends the message until the acknowledgement is received. The vehicles can send a~\emph{Change-Request} message to the manager if they need to update information in their previous request. Once a vehicle has finished crossing the conflict zone, it is required to send a~\emph{Done} message to the manager, which lets the manager know it can remove the corresponding vehicle's platoon from the schedule.

This protocol is designed to hide the scheduling policy from the vehicle agents, therefore allowing for online changes to the policy without requiring any changes in communications.
\subsection{Policy}
The scheduling problem is formulated as the following minimization problem.
\begin{equation}
    \label{eq:argmin}
    arg\min_s \delta(s) = \{s \mid s \in S \wedge \forall s' \in S: \delta(s) \leq \delta(s')\}
\end{equation}
Where $s$ is a schedule of vehicles (platoons), $\delta$ is a cost function designed to penalize total delay and variance in delay, given a schedule, as in Equations~\ref{eq:cost1} and~\ref{eq:cost2}. 
\begin{align}
    \delta_{1}(s) &= \sum_{j=2}^{N} j\left(d(p_{j}) + \sum_{i=1}^{j-1}t_{c}(p_{i})\right)\label{eq:cost1}\\
    \delta_{2}(s) &= \sum_{j=2}^{N} \left(d(p_{j}) + \sum_{i=1}^{j-1}t_{c}(p_{i})\right)\label{eq:cost2}
\end{align}
In the above equations,~$N$ is the number of platoons in the schedule,~$j$ is the platoon's turn,~$d$ is a function that returns the sum of the current delay of the vehicles within a given platoon, and again $s$ is the given schedule. This delay,~$d$, is the difference between the vehicle's original expected arrival time and its expected arrival time assuming it will be the next platoon going through.~$t_{c}$ is a function that computes the additional delay caused by the platoons that are given higher priorities than the~$j$th platoon in the schedule.

The scheduler in Equation~\ref{eq:argmin} essentially simulates all possible schedules given the set of platoons in the queue, and returns a schedule that has the lowest score in terms of the cost function~$\delta_{1}(s)$ or~$\delta_{2}(s)$. For the rest of this paper we will refer to~$\delta_{1}(s)$ and~$\delta_{2}(s)$ as~\emph{Platoon-based Variance Minimization} (\emph{PVM}) and \emph{Platoon-based Delay Minimization} (\emph{PDM}), respectively.

The scheduling procedure will be called by the controller every time a change is detected in the set of platoons in the schedule. For example, the procedure is called when a platoon finishes crossing the conflict zone and as a result is removed from the schedule, or when a new platoon enters the communication zone. The controller keeps a record of the most recent schedule and sends a message to the platoon that is at the top of the queue. The controller also checks for non-conflicting turning movements in the schedule with that of the platoon at the top of the queue, and lets them cross the intersection simultaneously. The clearance time of the intersection is then updated by the maximum estimated clearance time of the platoons that are crossing the conflict zone. 

Algorithm~\ref{alg:1} shows the intersection management algorithm in pseudo code.
\begin{algorithm}[h]

  \noindent\begin{minipage}{\linewidth}
   \caption{Autonomous Intersection Manager}
    \begin{algorithmic}[1]
    \Function{IntersectionManager}{}
	\While{$True$}
		\State $P = getRequests()$ \footnote{Where P is a map of platoons paired with their respective request information}
		\State $sort(P)$\footnote{Sort the platoon list based on their expected arrival time to make the candidate selection run faster}
		\State $pool = selectCandidates(P)$
		\If{$!pool.isUpdated()$}
			\State $continue$ \footnote{Skip this iteration if the selection pool has not changed}
	    \EndIf
	    \State $[platoons,schedule] = getSchedule(pool)$
		\State $i=1$
        \For{$platoon$ in $platoons$}
        	\State $update(platoon,schedule_{i})$
        	\State $i++$
        \EndFor
        \EndWhile
       \EndFunction
\end{algorithmic}
    \label{alg:1}
\end{minipage}
\end{algorithm}

The algorithm considers at most one platoon for each lane that is in the communication range of the central controller, i.e. the leading platoon in each lane. Such design makes the policy scalable, in that the computational complexity of the scheduler remains suitably low as the number of incoming lanes grow.

\subsection{Computational Complexity}
The algorithm considers all permutations of the set of platoons, including possible non-conflicting trajectories in which case simultaneous crossing is considered. These permutations can be modeled as a permutation problem to pick from~$K$ elements without replacement and placing them in sets of~$\{K, K-1, ..., 1\}$ placeholders. Therefore, the worst case computational complexity of such algorithm would be equal to the number of possible permutations, given in Equation~\ref{eq:perms}.
\begin{equation}
    \label{eq:perms}
    T = \sum_{r=1}^{N}\sum_{i=0}^{r-1} (-1)^{i} {r\choose i}(r-i)^N
\end{equation}

In the above equation, $T$ is the total number of possible schedules, $N$ is the number of lanes, $r$ and $i$ are index counters. One may note that the exponential nature of this complexity can only be acceptable for small number of incoming lanes. It can be shown that the algorithm will have the worst case computational complexity of $O(N^N)$. For example, the 4-way intersection considered in this work would only require 75 possible schedules to be considered in the worst case. Due to space constraints, the details and proof of computational complexity are omitted.

\begin{figure}[H]
    \vspace{-11pt}
    \centering
    \includegraphics[width=.95\columnwidth]{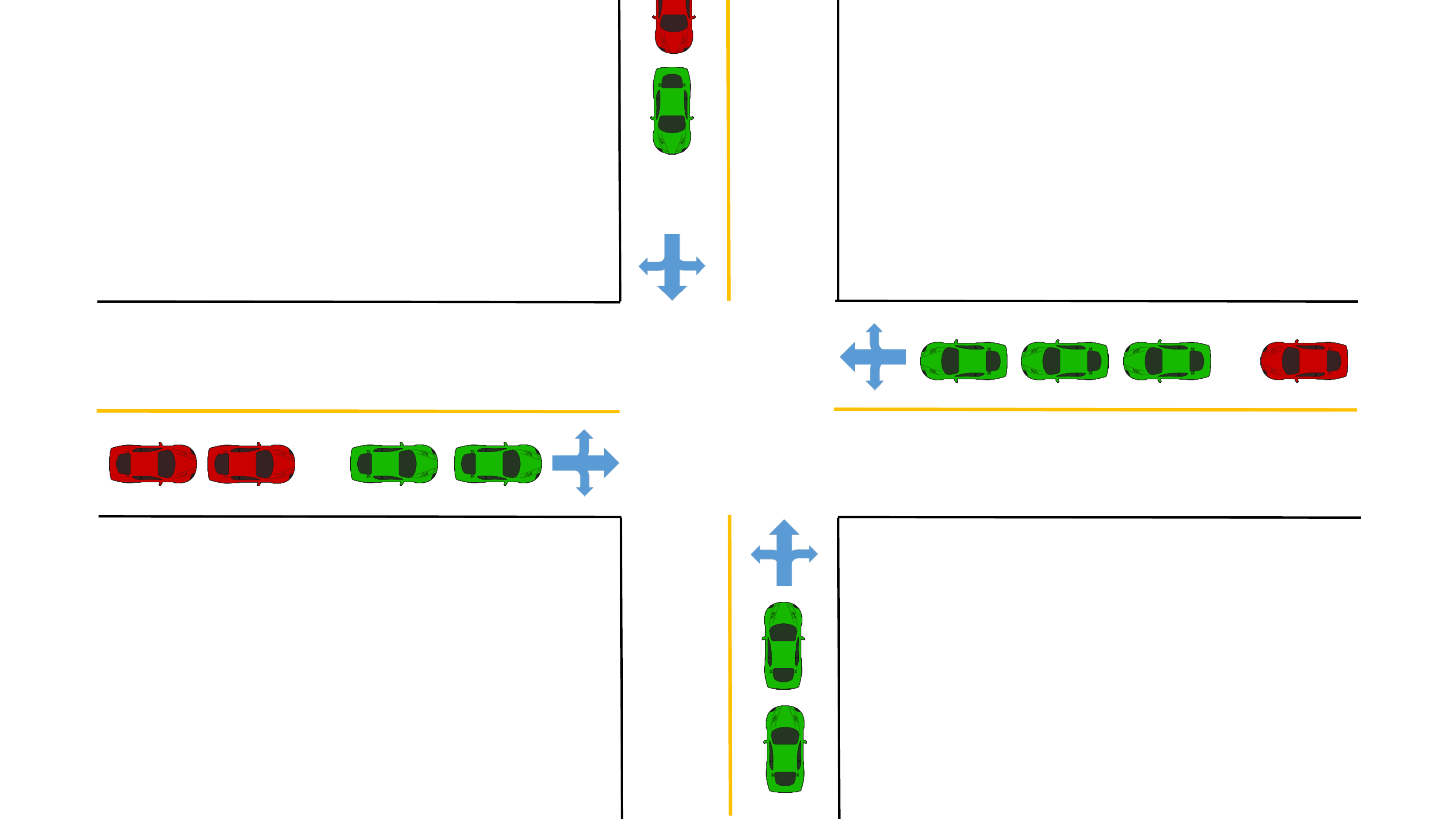}
    \caption{4-way Intersection Geometry}
    \label{fig:intersection}
\end{figure}

We propose a heuristic for larger intersections to reduce computational complexity. The proposed heuristic does not consider non-conflicting trajectories and therefore reduces the complexity to the number of possible permutations which is exactly $N!$, i.e. computational complexity of $O(N!)$. After a schedule is selected, the controller allows those platoons with non-conflicting trajectories with regard to the selected platoon to cross the intersection simultaneously. Fig.~\ref{fig:intersection} demonstrates how the proposed heuristic ignores the platoons behind the closest platoon to reduce the computational complexity. The platoons/cars in red represent the ignored input to the algorithm. This figure also serves as a visual aid to represent the geometry and turning policy of the intersection that was used for the simulations.

\section{Simulations}\label{sec:simulations}
\begin{figure}[bh]
    \centering
    \includegraphics[width=.95\columnwidth,trim={0 1.5cm 0 1.5cm},clip]{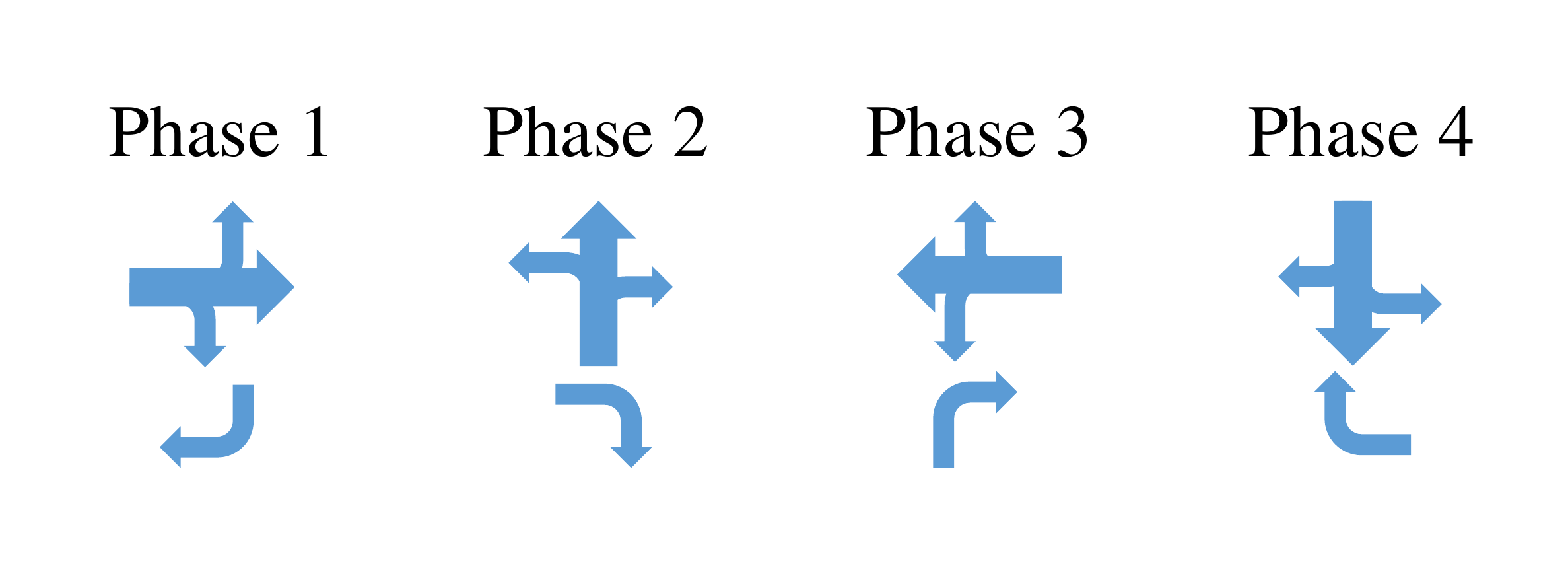}
    \caption{Phase Plan}
    \label{fig:phaseplan}
\end{figure}

\begin{figure}[th]
    \centering
    \includegraphics[width=\columnwidth,trim={0 0.5cm 0 0.5cm},clip]{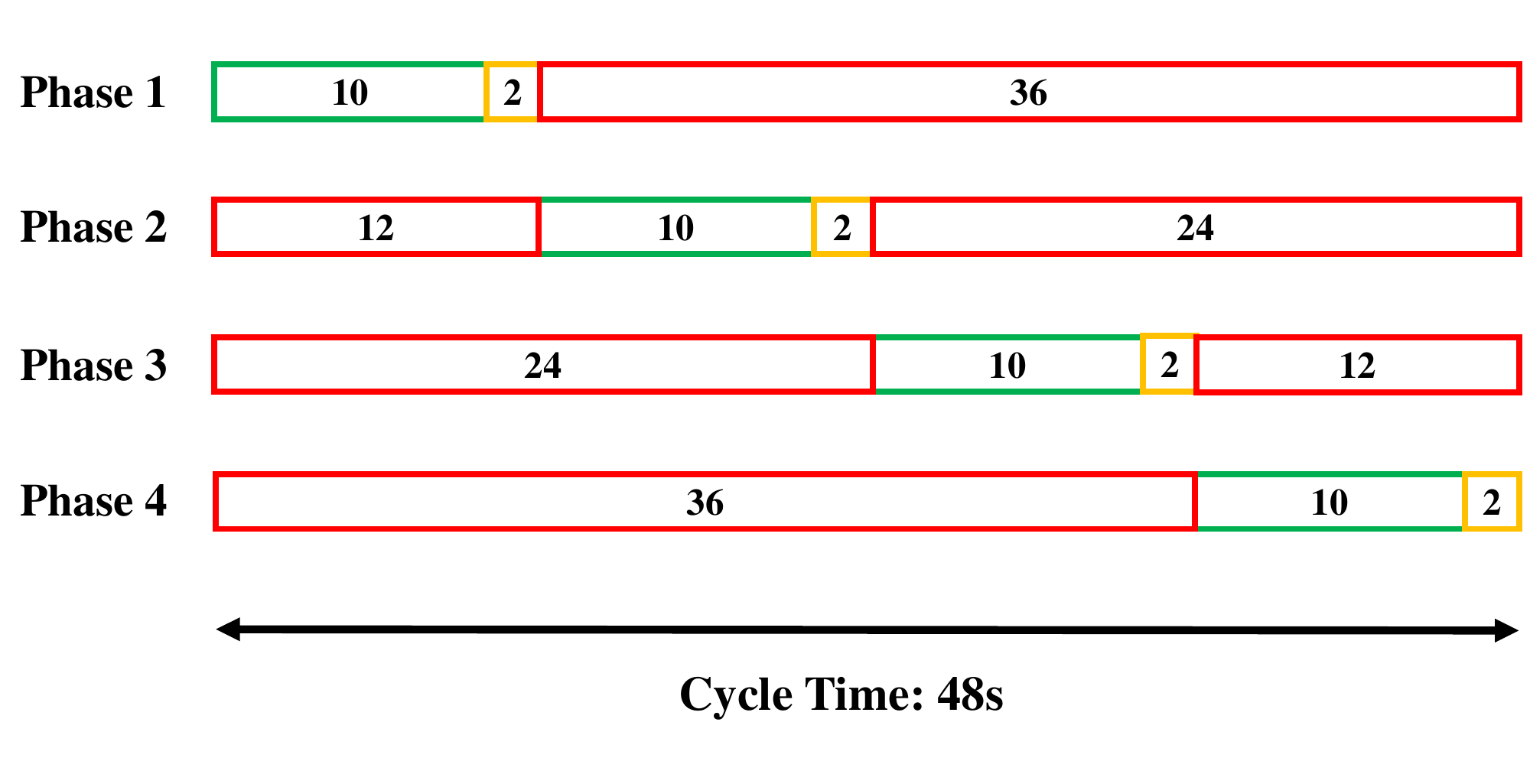}
    \caption{Timing Diagram}
    \label{fig:timing}
    \vspace{-0.6cm}
\end{figure}

A fixed-time 4-phase traffic light controller is tuned for the 4-way intersection shown in Fig.~\ref{fig:intersection}. The phase plan and timing diagram of the baseline traffic light policy are shown in Figs~ \ref{fig:phaseplan} and~\ref{fig:timing} respectively.

\begin{figure*}
\centering
\begin{minipage}[H]{.45\textwidth}\centering
      \includegraphics[scale=0.31]{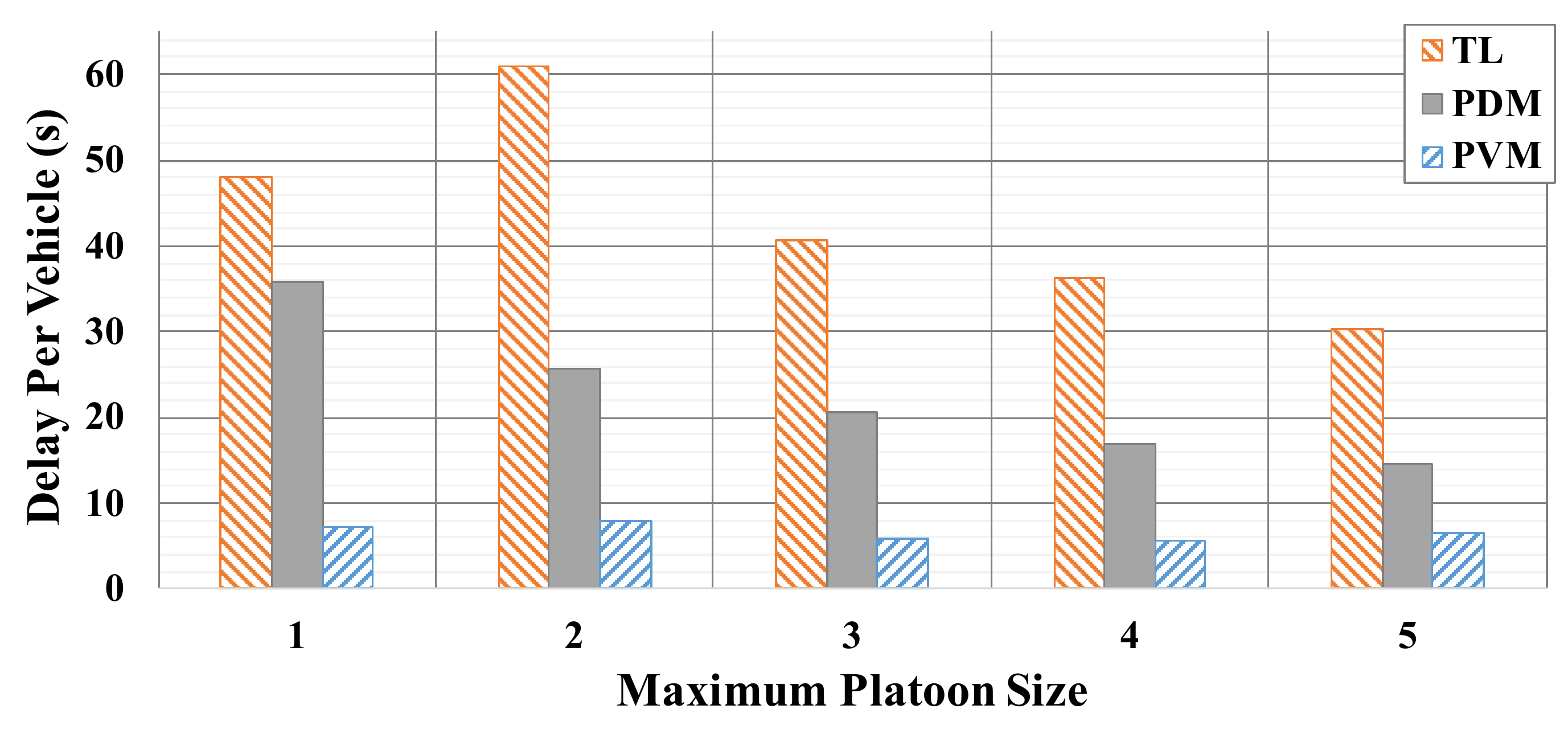}
\caption{Average Delay Per Vehicle}\label{fig:1}
      \vspace{-0pt}
\end{minipage}\qquad
\begin{minipage}[H]{.45\textwidth}\centering
      \includegraphics[scale=0.31]{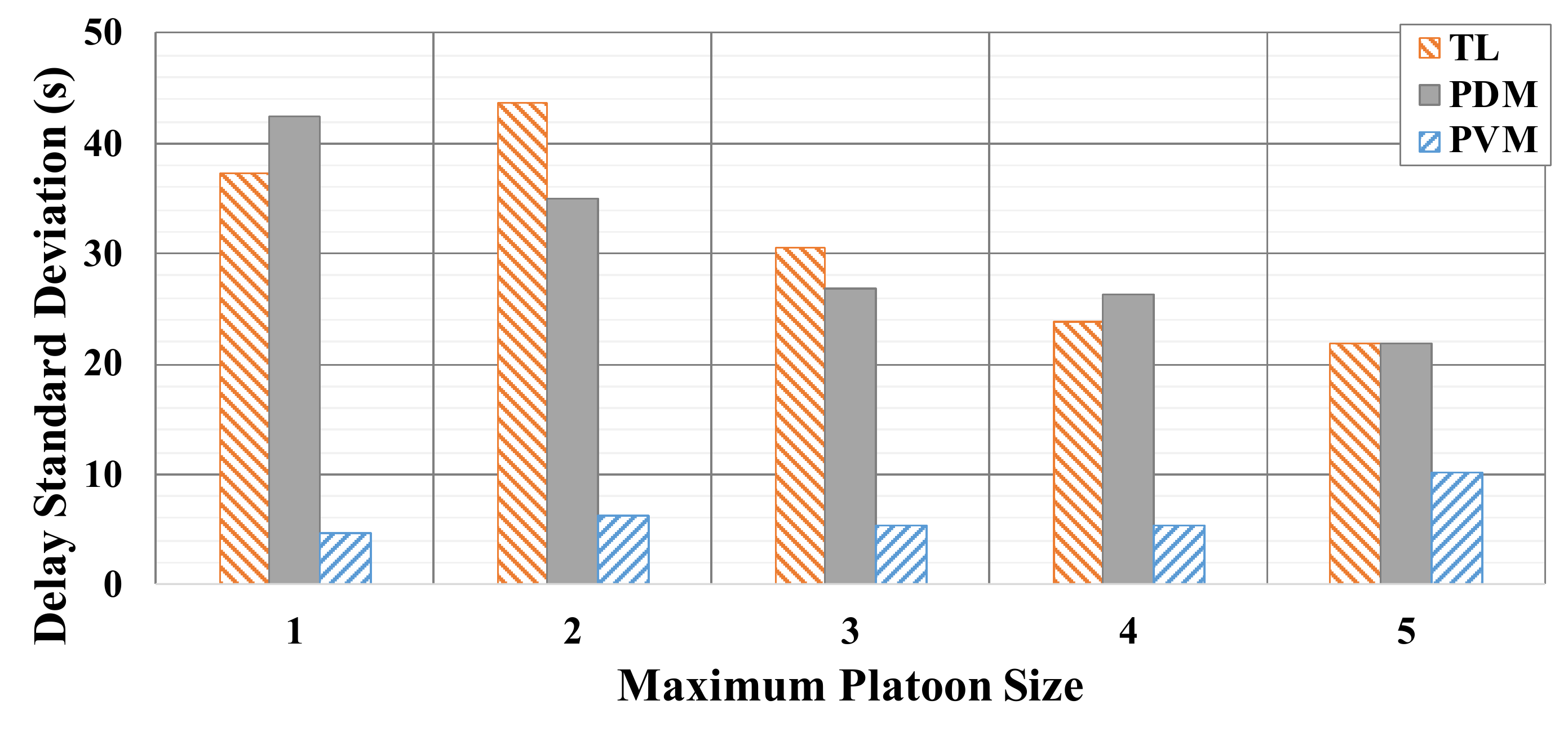}
\caption{Delay Standard Deviation}\label{fig:2}
      \vspace{-0pt}
\end{minipage}
\begin{minipage}[H]{.45\textwidth}\centering
      \includegraphics[scale=0.31]{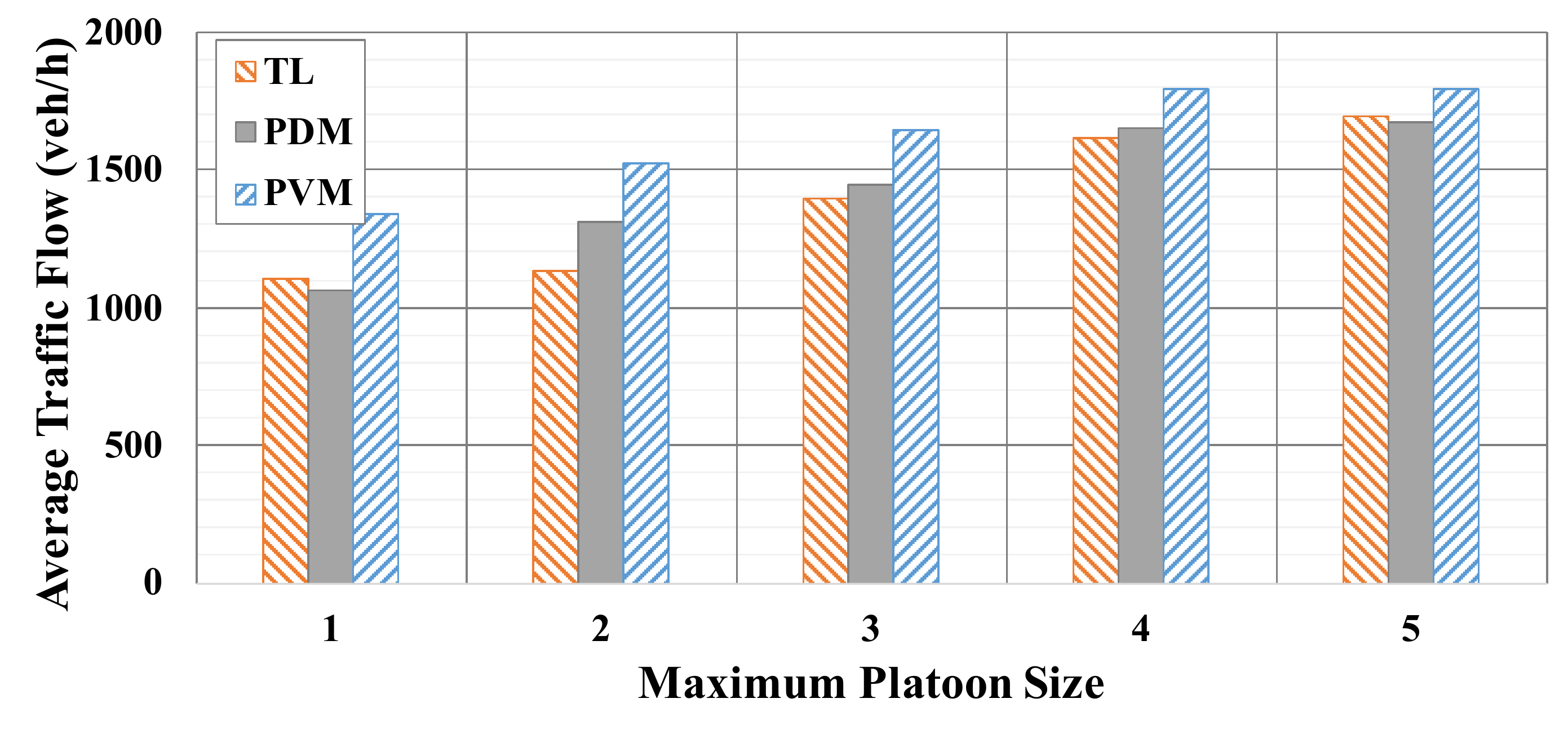}
\caption{Intersection Traffic Flow}\label{fig:3}
      \vspace{-8pt}
\end{minipage}\qquad
\begin{minipage}[H]{.45\textwidth}\centering
      \includegraphics[scale=0.31]{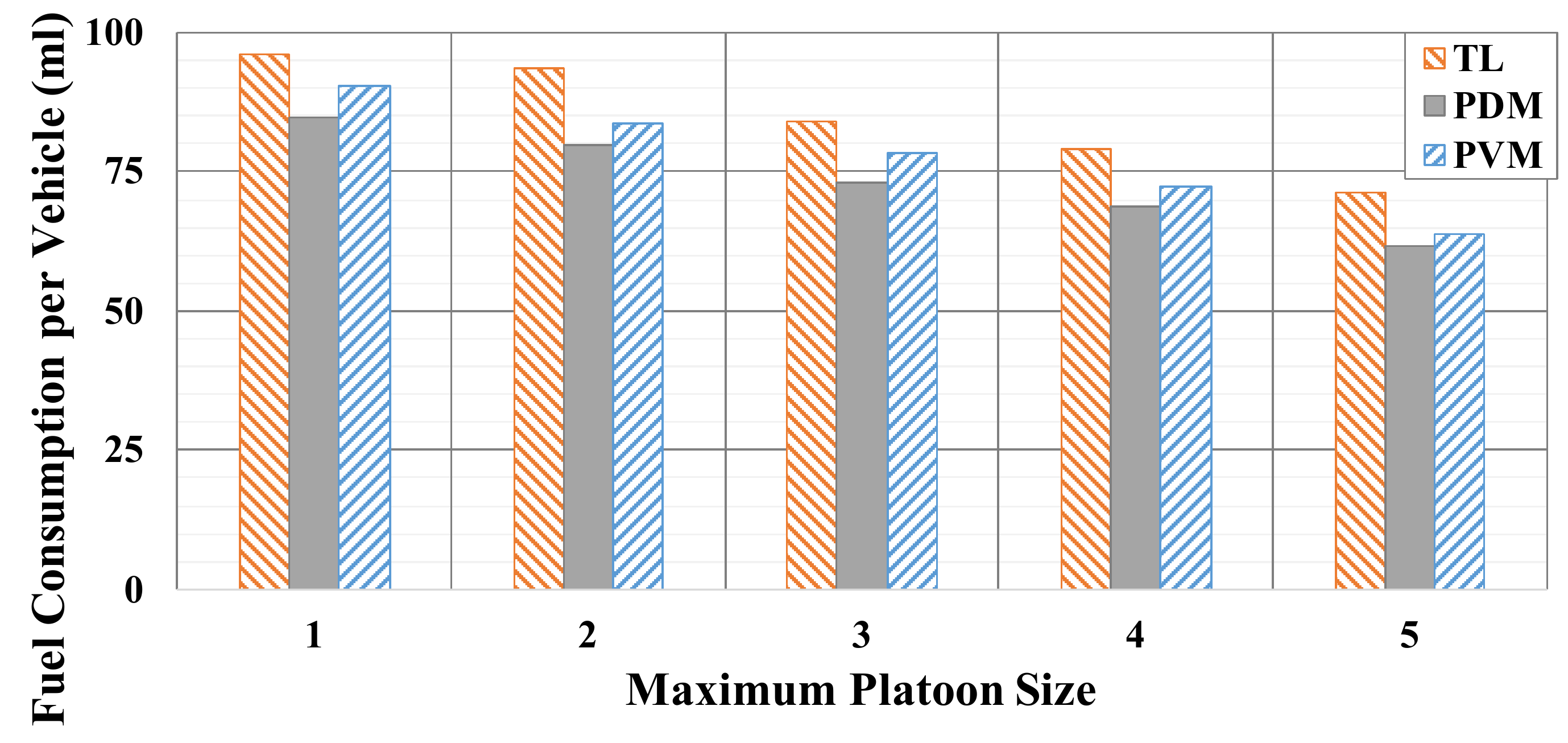}
\caption{Fuel Consumption}\label{fig:4}
      \vspace{-8pt}
\end{minipage}
\end{figure*}

To compare the performance of the two proposed methods against the baseline policy, we design 20 scenarios by choosing different parameter settings in terms of incoming traffic flow and maximum platoon size. Each policy is evaluated for every variation of the two parameters in table~\ref{tab:parameters}. The incoming traffic flow is equal on all approaches with $70\%$ of the traffic going straight,~$20\%$ turning right and~$10\%$ turning left. To minimize the influence of randomness, each simulation is run for 60 minutes.
\begin{table}[h]
\centering
\caption{Simulation Parameters}
\label{tab:parameters}
\begin{tabular}{@{}ccc@{}}

                  Parameter Name   & Set of Values       & unit    \\ \midrule
Traffic Level        & \{500,600,700,800\} & veh/hour/lane \\ \midrule
Simulation Time        & \{3600\} & s \\ \midrule
Maximum Platoon Size & \{1,2,3,4,5\}       & veh     \\ \bottomrule
\end{tabular}
\end{table}
Recorded videos of simulation for~\emph{PVM}\footnote{\href{https://youtu.be/RtN0f7BlFyg}{https://youtu.be/RtN0f7BlFyg}} and~\emph{PDM}\footnote{\href{https://youtu.be/qHGv9LF72NA}{https://youtu.be/qHGv9LF72NA}} are available. Figs~\ref{fig:1},~\ref{fig:2},~\ref{fig:3} and~\ref{fig:4} demonstrate the results in terms of delay per vehicle, delay standard deviation, intersection capacity and fuel consumption, respectively. To conserve space and promote readability, the results are aggregated over the set of values for the traffic level parameter.

According to Fig.~\ref{fig:1}, the proposed methods significantly outperform the traffic light controller in terms of average delay per vehicle. Fig.~\ref{fig:2} shows the computed standard deviation of delays throughout the entire simulations for each policy and the set of maximum platoon sizes. It can be seen that the~\emph{PVM} method significantly decreases the standard deviation compared to the traffic light, while as expected,~\emph{PDM} does not show a meaningful improvement in terms of delay variance as its cost function is designed to solely decrease total delay. One may also note that the maximum platoon size and the~\emph{PVM} and the traffic light performance are negatively correlated, which confirms the positive effect of platooning on the performance of any type of traffic controller.

Fig.~\ref{fig:3} shows that platoon size and intersection capacity are positively correlated for all policies. For the traffic light policy, larger platoons result in more smooth trajectories with shorter headways, and as a result increases the outgoing traffic flow. Larger platoons also help the proposed policies select better schedules in terms of total delay and variance, which ultimately will increase the outgoing traffic flow. In the simulations, \emph{PVM} policy outperformed traffic light for all incoming traffic flows and platoon sizes.

Fig.~\ref{fig:4} demonstrates fuel consumption per vehicle as a function of platoon size. As expected, platoon size and fuel consumption are strongly correlated. \emph{PVM} and~\emph{PDM} policies outperform traffic light by~$8\%$ and~$13\%$ on average, respectively. This result can be explained by the shorter idle times generated by~\emph{PVM} and~\emph{PDM} compared to the traffic light policy. According to the fuel consumption model adopted in the simulations, the vehicles consume fuel at a rate of~$0.15~ml/s$ when idle.
\begin{table*}[h]
\caption{Aggregated Results}
\label{tab:final}
\centering
\begin{tabular}{p{0.15\linewidth}p{0.10\linewidth}p{0.10\linewidth}p{0.10\linewidth}p{0.15\linewidth}p{0.15\linewidth}}
\hline
 & Traffic Light & PVM               & PDM      & PVM-Improvement & PDM-Improvement \\
 \hline
Delay(s)              & $43.26$       & $\bm{6.56}$   & $22.71$  & $\bm{6.6\times}$     & $1.9\times$    \\ 
Capacity(veh/h)        & $1388$      & $\bm{1617}$ & $1426$ & $\bm{13.8\%}$      & $2.7\%$ \\ 
FCPV(ml/v)             & $84$      & $77$ & $\bm{73}$ & $8\%$ & $\bm{13\%}$      \\ 
STDEV(s) & $31.44$         & $\bm{6.37}$     & $30.48$    & $\bm{4.9\times}$ & $3\%$  \\    
\hline
\end{tabular}
\vspace{-11pt}
\end{table*}

To compare the overall performance of the policies, results from all configurations are aggregated into table~\ref{tab:final}. All the metrics in this table are averaged over the set of incoming traffic flows that range from~$500$ to~$800~v/h/l$. Traffic flows are identical for each approach.

The~\emph{PVM} method outperforms the traffic light policy on all four metrics. More notably, it decreased average delay per vehicle by factor of~$6.56\times$ and decreases the standard deviation to~$4.9\times$, resulting in faster and more reliable traffic flows. The~\emph{PVM} policy also increased the intersection capacity by~$13.8\%$ compared to traffic light.

The~\emph{PVM} and~\emph{PDM} policies both outperform the traffic light in terms of fuel consumption by~$8\%$ and~$13\%$ respectively.

\section{Conclusions}\label{sec:conclusions}
In this paper, a centralized platoon-based controller was proposed for the cooperative intersection management problem that takes advantage of the platooning systems to generate fast and smooth traffic flow at an intersection. A simple communication protocol was designed for V2I communication and two policies were introduced for the controller to minimize total delay and delay variance according to the cost functions tailored for platoons of vehicles.

According to the simulation results, the proposed controller minimizes travel delay and variance while increasing intersection throughput and reducing fuel consumption, when compared to traffic light policies. The simulations also verify the positive effect of platoon size on fuel consumption and intersection throughput.


%




\ifCLASSOPTIONcaptionsoff
  \newpage
\fi



\bibliographystyle{IEEEtran}
%

\bibliography{references}

%





\end{document}